# On the Quantum-Optical Nature of High Harmonic Generation

**Alexey Gorlach[1], Ofer Neufeld[1], Nicholas Rivera[2], Oren Cohen[1], and Ido Kaminer[1]**

[1] Technion – Israel Institute of Technology, 3200003 Haifa, Israel

[2] Massachusetts Institute of Technology, 02139 Cambridge MA, USA

kaminer@technion.ac.il

**High harmonic generation (HHG) is an extremely nonlinear effect, where a medium is driven by a strong laser field, generating coherent broadband radiation with photon energies ranging up to the X-ray and pulse durations reaching attosecond timescales. Conventional models of HHG treat the medium quantum mechanically, while the driving and emitted fields are treated classically. Such models are usually very successful, but inherently cannot capture the quantum-optical nature of the process. Despite prior works considering quantum HHG, it is still not known in what circumstances the spectral and statistical properties of the radiation considerably depart from the known phenomenology of HHG. Finding such regimes in HHG could lead to novel sources of attosecond light with intrinsically quantum statistics such as squeezing and entanglement. In this work, we present a fully quantum electrodynamical theory of extreme nonlinear optics with which we predict new quantum effects that alter both spectral and statistical properties of HHG. We predict the emission of shifted frequency combs resulting from transitions between perturbed states of the driven atom, and identify new spectral features that arise from the breakdown of the dipole approximation for the emission. Moreover, we find that each HHG emitted photon is a superposition of all frequencies in the spectrum – e.g., HHG creates single photon combs. We also describe how the HHG process changes in the single atom regime, and discuss how our various predictions can be tested experimentally. Our approach is also applicable to a wide range of nonlinear optical processes, paving the way toward novel quantum information techniques in the EUV and X-ray, and in ultrafast quantum optics.**

**Section I – Introduction**

High-harmonic generation (HHG) is a physical effect that occurs when an atomic, molecular, or solid system is placed in a strong driving laser field and emits photons at frequencies of integer multiples of the driving field frequency [1, 2]. HHG provides a coherent source of extreme ultraviolet (XUV) emission and has also paved the way to the field of attoscience [3, 4]. This intriguing process has been under investigation for the last several decades, and it is standardly well-described by the so-called three-step model [2, 5]. According to this model: (i) the electron tunnels out from the atomic potential suppressed by the intense driving field, (ii) is consequently accelerated in the continuum by the driving field, (iii) and under certain conditions can return to the ion and recombine, thus emitting a high energy photon. This process repeats itself periodically, resulting in a comb emission. A better quantitative understanding of the phenomena of HHG was made by the highly successful semi-analytical quantum theory by Lewenstein [6], where the electron is described quantum mechanically and the driving and emitted fields are still described classically. Many advances in the theory have followed since, particularly regarding accurate *ab-initio* treatments for the HHG process from atoms and molecules [7, 8, 9, 10], as well as the description of various HHG mechanisms from solids [11, 12, 13, 14].

All such theories describe the dynamics of the electrons in the driving field quantum mechanically, however, treat the emission classically, as dipole radiation. The dipole sources are calculated as the expectation values of the dipole moment of each driven atom [15]. Some early pioneering theoretical works [16, 17], as well as some more recent works [18, 19], have developed a quantum electrodynamical formalism to describe HHG by quantizing the emitted field. The driving field was also quantized in some recent experimental and theoretical works [20, 21, 22, 23, 24]. However, it is still not known in what conditions the spectral and statistical properties of the radiation differs significantly from the known effects of HHG, as currently seen in experiments. It remains an open question whether the quantum theory can bring new effects of intrinsically quantum nature, such as non-classical photon statistics or entanglement of the emission and emitting media. In particular, it remains unknown whether the emission should be thought of as an ensemble of photons, each with a single frequency (in a mixed state)? Or is each photon a quantum superposition of all frequencies in the comb? Answering these questions would reveal new aspects of HHG with implications in both attoscience, quantum optics, quantum electrodynamics (QED), and quantum information.

Here we analytically develop a full quantum theory of extreme nonlinear optics and use it to explore the quantum-optical nature of HHG. Our formalism does not assume a specific electronic system – it applies for atoms, molecules, or solids – we use the term "atom" in the sense of a general system. We present predictions both for HHG in the single-atom regime, as well as HHG from many (an ensemble of) atoms, and highlight in each case the deviation from the conventional treatments. In particular, in the single-atom case, we show that the spectrum would contain multiple shifted combs of HHG, which arise due to transitions between initial and different final states of the driven atom. For both a single and many atoms, we find new features in the HHG spectrum arising from the breakdown of the dipole approximation to the emitted photons, which can be observed as the emission of even harmonics (even from a monochromatic driving field). Most importantly, we show that each HHG emitted photon is a comb with attosecond timescales and carries the entire spectrum's spectral content, which may be measured by a field autocorrelation experiment. Consequently, even a single photon carries information about the HHG process, including the energy destitution and the cut-off frequency, up until it is observed.

The paper is organized as follows: in Section II we develop a fully quantum formalism of extreme nonlinear optics using strong-field quantum electrodynamics (SFQED) and apply it for HHG; in Section III we show the existence of shifted frequency combs in the single atom regime and discuss the differences in the many atom regime; in Section IV we find HHG spectral corrections that arise from the breakdown of the dipole approximation; in Section V we propose quantum-optical signatures of the HHG emission, and discuss an experimental test.

**Section II – Quantum theory of extreme nonlinear optics and high-harmonic generation**

In this section, we develop a general, fully quantum framework for predicting the emission from an electronic system in a strong, time-dependent external electromagnetic field. We consider an electronic system driven by a strong laser field, which is described by a multimode coherent state: $|\psi_{\text{laser}}\rangle = \prod_{k\sigma}|\alpha_{k\sigma}e^{-i\omega_k t}\rangle$, where $\alpha_{k\sigma}$ are the coherent states' parameters that can be shown to be equal to the complex amplitudes of the Fourier components of the classical description of the incident driving field, $\boldsymbol{k}$ refers to the wavevector of a plane wave in free space, $\sigma$ refers to its polarization, and $\omega_k = ck = c|\boldsymbol{k}|$ to its frequency, with $c$ the speed of light in vacuum. The combined wavefunction of the electronic system and the electromagnetic field, $|\Psi(t)\rangle$, is determined by the Schrodinger equation:

$$i\hbar \frac{\partial}{\partial t}|\Psi(t)\rangle = H|\Psi(t)\rangle, \qquad (1)$$

where the (QED) Hamiltonian in the case of one electron is $H = \frac{1}{2m}(\boldsymbol{p} - q\boldsymbol{A})^2 + U + H_F$, with $q$ being the electron charge and $m$ being and the electron mass, $U$ the atomic potential, and $H_F$ the Hamiltonian of the free electromagnetic field. The quantized vector potential $\boldsymbol{A}$ contains both the driving field and the emitted field. This Hamiltonian is only brought as an example, and the entire formalism below can be applied for any Hamiltonian. The solution of Eq. (1) for the combined wavefunction $|\Psi(t)\rangle$ relies on three important steps.

In the first step, we perform a unitary transformation on the Hamiltonian, which decomposes the entire vector potential $\boldsymbol{A}$ into the sum of a classical time-dependent part $\boldsymbol{A}_c(t) \triangleq \langle\psi_{\text{laser}}(t)|\boldsymbol{A}|\psi_{\text{laser}}(t)\rangle$ and a small quantum correction $A_q \triangleq \sum_{k\sigma} \sqrt{\frac{\hbar}{2\varepsilon_0 Vck}} e_\sigma [a_{k\sigma} e^{ik \cdot r} + a_{k\sigma}^\dagger e^{-ik \cdot r}]$. Here $V$ is the volume of space, $\sum_{k\sigma}$ a summation over all possible photonic modes, $a_{k\sigma}$ and $a_{k\sigma}^\dagger$ are the annihilation and creation operators of a photon with polarization $\sigma$ and wavevector $k$, $e_\sigma$ is a unit vector of polarization, and $\varepsilon_0$ is the vacuum permittivity. $A_q$ describes the quantum emitted field, and $A_c(t)$ describes the classical driving laser field.

<u>In the second step</u>, we can take advantage of existing analytical and numerical techniques that have been widely developed to solve the time-dependent Schrodinger equation (TDSE) [15, 25]:

$$i\hbar \frac{\partial |\phi_i(t)\rangle}{\partial t} = H_{\text{TDSE}} |\phi_i(t)\rangle, \qquad (2)$$

where $|\phi_i(t)\rangle$ is the wavefunction of the electronic system, whose state at an initial time ($t = -\infty$) is typically the ground state or some other eigenstate of the electronic system, but can also be a superposition of eigenstates. The TDSE Hamiltonian $H_{\text{TDSE}}$ depends on the electronic system and for a single electron has the form $H_{\text{TDSE}} = \frac{1}{2m}(p - qA_c(t))^2 + U + H_F$.

<u>In the third step</u>, we calculate the quantized radiation emission by the interaction of $A_q$ with the strongly driven electronic system state $|\phi_i(t)\rangle$ calculated in the second step. As shown in Supplementary Material (SM), the coupling between an electronic system and the emitted field in typical nonlinear optics experiments such as HHG is weak, and therefore can be accounted for by perturbation theory. The combined wavefunction $|\Psi(t)\rangle$ of the driven atomic system and the photons is constructed from two parts: (1) Each time-dependent electronic wavefunction $|\phi_i(t)\rangle$ is calculated using the classical driving field $A_c(t)$. (2) Each photon emission is calculated to first order in the quantized field $A_q$. Together, these two parts show how to calculate first-order processes in SFQED. This situation is shown in the left column of Figure 1.

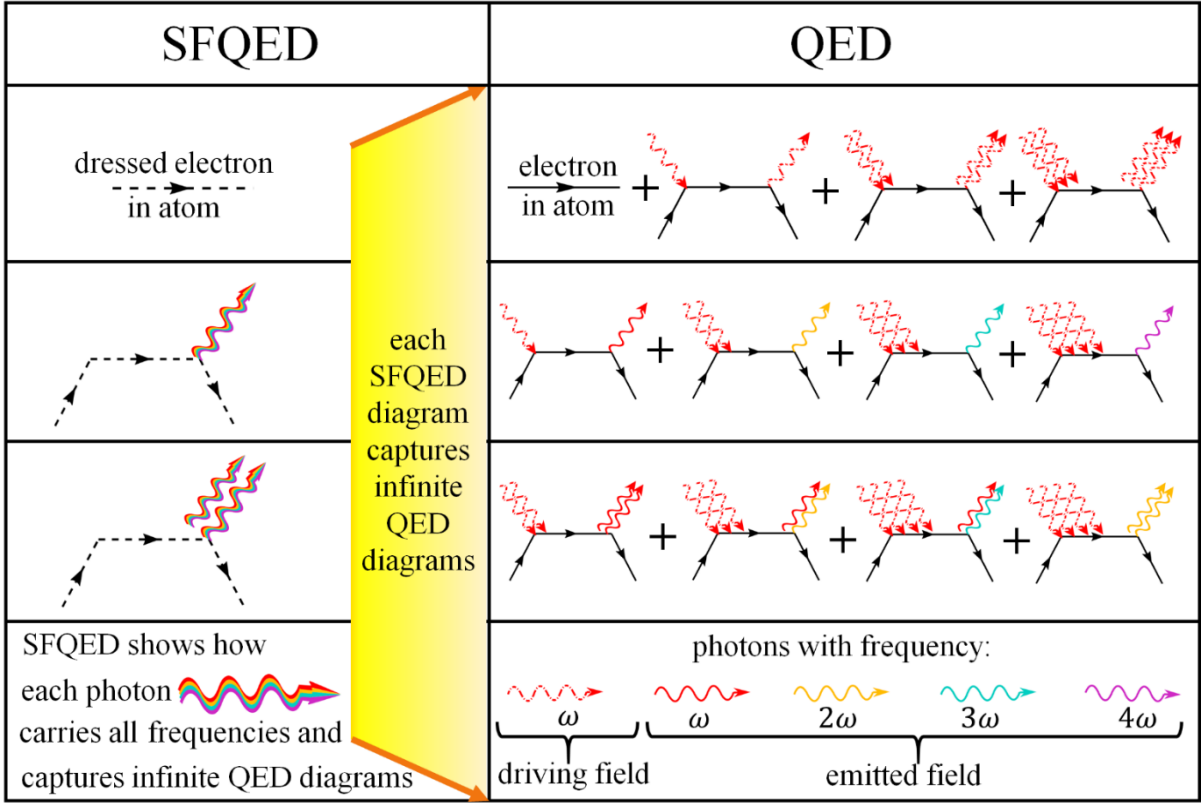

**Figure 1: Diagrams in strong-field quantum electrodynamics (SFQED) versus ordinary quantum electrodynamics (QED).** In SFQED, we have perturbation theory in the weak emitted field: different rows correspond to a different number of emitted photons. Each perturbation order in SFQED is equivalent to infinite diagrams of emission and absorption processes in the standard formulation of QED. The first row corresponds to zero-order SFQED, containing all the QED diagrams necessary to describe the evolution of the wavefunction according to the TDSE (Eq. (2)). The second row corresponds to first-order SFQED, which we use to capture the effect of HHG. The third row corresponds to second-order SFQED, which can capture new processes in extreme nonlinear optics such as laser-

In the framework of SFQED, frequency conversion in nonlinear optical effects (such as HHG) is equivalent to a process of spontaneous emission with transitions between the electronic states $|\phi_i(t)\rangle$ that are dressed by the laser field [26]. The time-dependent state $|\phi_i(t)\rangle|0\rangle$, with $|0\rangle$ being a state with no emitted photons, makes a quantum mechanical transition into states of the form $|\phi_j(t)\rangle|1_{k\sigma}\rangle$. The time-dependent states, $|\phi_j(t)\rangle$, all evolve according to Eq. (2) and form a basis for the electronic system. The photonic state $|1_{k\sigma}\rangle$ represents one photon with wavevector $\boldsymbol{k}$ and polarization $\sigma$.

In the rest of this section, we demonstrate the results of our formalism for a general nonlinear optical process – the resulting formulas also directly describe HHG. We use the dipole approximation to show explicit analytical expressions for the combined wavefunction $|\Psi(t)\rangle$ and for the emission spectrum. Section IV shows corrections to the emission spectrum beyond the dipole approximation. We relegate technical details to the SM. Under such conditions $|\Psi(t)\rangle$ is given by:

$$|\Psi(t)\rangle = \left(|\phi_\text{i}(t)\rangle|0\rangle + \frac{1}{\hbar}\sum_j \sum_{\boldsymbol{k}\sigma} e^{-i\omega_k t}\sqrt{\frac{\hbar\omega_k}{2V\varepsilon_0}}\left[\int_{-\infty}^{t} d\tau\, \left(\boldsymbol{d}_{ji}(\tau)\cdot e^{i\omega_k \tau}\boldsymbol{e}_\sigma\right)\right]|\phi_j(t)\rangle|\boldsymbol{k}\sigma\rangle\right), \quad (3)$$

where $\boldsymbol{d}_{ji}(\tau) = \langle\phi_j(\tau)|q\boldsymbol{r}|\phi_\text{i}(\tau)\rangle$ is a dipole matrix element that can be complex. The summation $\sum_j$ is performed over all possible final electronic states $|\phi_j(t)\rangle$, labeled by different quantum numbers $j$. Each final (time-dependent) electronic state corresponds to a photonic state that is a superposition of various single-photon states with wavevector $\boldsymbol{k}$ and polarization $\sigma$. The result for $|\Psi(t)\rangle$ is independent of the choice of the basis $|\phi_j(t)\rangle$.

Importantly, the superposition state in Eq. (3) shows that any emission process in nonlinear optics in which the electronic system varies in time does not emit photons with different fixed frequencies, but rather, each photon can be a superposition of multiple frequencies. This conclusion applies directly to HHG, showing that each photon carries the entire HHG spectrum, containing all the spectral information of the HHG pulse (e.g., a single photon comb), which is a key conclusion that is illuminated by our formalism. The remarkable feature of the combined wavefunction in Eq. (3) is the *entanglement* between the photonic state and the electronic state, in the sense that we cannot decompose the wavefunction to be a tensor product of photonic and electronic states. The entanglement implies that there remain a connection between the photon and the emitting atom after emission, which may have intriguing consequences in the field of quantum optics.

Our SFQED formalism yields predictions such as the emitted photon energy per unit frequency $\frac{d\varepsilon}{d\omega}$ (i.e., the spectrum), which follows immediately from modulo-squaring the final-state amplitudes of Eq. (3) and integrating over all photon emission angles and polarizations (see SM). The spectrum is given by:

$$\frac{d\varepsilon}{d\omega} = \sum_j \frac{\omega^4}{6\pi^2\varepsilon_0 c^3} \left| \int_{-\infty}^{+\infty} \boldsymbol{d}_{ji}(t) e^{i\omega t} dt \right|^2. \tag{4}$$

We can find $\boldsymbol{d}_{ji}(t)$ by solving the TDSE for each $j$. When solving the TDSE for a single electron in a single atom, Eq. (4) yields the general result for the spectrum of HHG from a single-atom. Numerical results using Eq. (4) are presented in Section III.

**Section III – Conceptual differences between single-atom HHG and many-atom HHG**

In this section, we numerically calculate the HHG spectrum from a single-atom in the 1D model of a helium atom. To emphasize the differences between the single-atom and many-atoms HHG, we compare our general result in Eq. (4) with the conventional formula of HHG [15], which has the following form:

$$\frac{d\varepsilon}{d\omega} = \frac{\omega^4}{6\pi^2\varepsilon_0 c^3} \left| \int_{-\infty}^{+\infty} \langle \boldsymbol{d}(t) \rangle e^{i\omega t} dt \right|^2, \tag{5}$$

where $\langle \boldsymbol{d}(t) \rangle \triangleq \boldsymbol{d}_{ii}(t) = \langle \phi_i(t) | q\boldsymbol{r} | \phi_i(t) \rangle$. Unlike the conventional Eq. (5) that contains only the expectation value of the dipole moment, our Eq. (4) contains all transition matrix elements $\boldsymbol{d}_{ji}$. We can always choose the basis of time-dependent states $|\phi_j(t)\rangle$ such that it includes $|\phi_i(t)\rangle$, and then identify the contributions from transitions involving electronic states other than the initial state as quantum corrections to the conventional prediction.

To see the conceptual differences between Eqs. (4) and (5), we performed then numerical calculation for a model of a helium atom depicted in Figure 2. We studied the dynamics of a helium atom (within the single active electron approximation) interacting with an external electric field with frequency $\omega$: $\boldsymbol{A}_c(t) = \frac{1}{\omega}\boldsymbol{E}_0 \cos \omega t$. The initial state of the atom was chosen as its ground state $|\phi_1\rangle$, mimicking a hydrogenic 1s state (depicted in Figure 2a). We calculate numerically the emission spectrum using Eqs. (4) and (5). As it is shown in Figure 2b, Eq. (4) gives much larger emission rates than the emission in Eq. (5). The spectrums of Eqs. (4) and (5) have such differences because transition matrix elements $\boldsymbol{d}_{ji}$ can be comparable to the element $\boldsymbol{d}_{ii}$ and even be much larger as depicted in Figure 2c,d,e,f. Moreover, from Figure 2b we can see that Eq. (5) gives a standard HHG spectrum with odd-only harmonics, while Eq. (4) has no distinguishable discrete harmonic peaks [27].

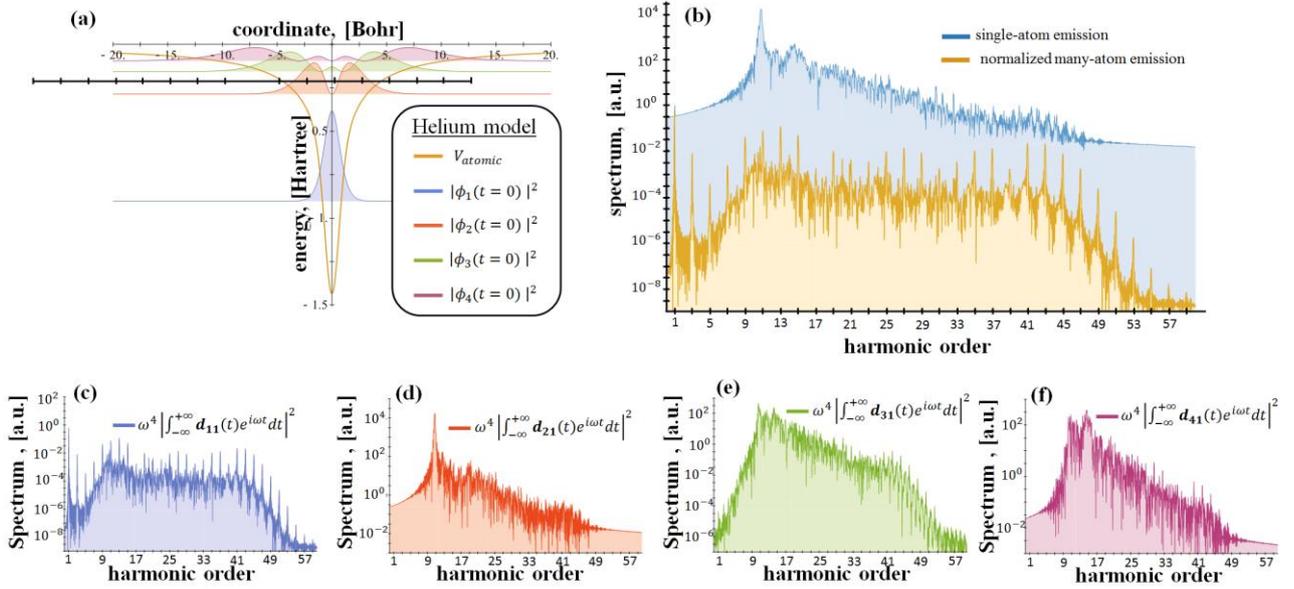

**Figure 2: Single-atom and many-atom HHG.** **(a)** 1D model of a helium atom and its electronic states ($|\phi_1\rangle$, $|\phi_2\rangle$, $|\phi_3\rangle$, etc.) without an external driving field. This model of a helium atom is used for all the simulations in our work. **(b)** Emission spectrum calculated using Eq. (5) (yellow curve) vs. summing the first four elements in Eq. (4) (blue curve), which present the HHG emission by a many ($N \gg 10^4$) atoms normalized by $N^2$ (yellow) and by a single atom (blue). **(c-f)** Contributions to the single-atom HHG spectrum, corresponding to different transition matrix elements, all having the same normalization. The wavelength of the driving field is $\lambda_0 = 800$nm; the intensity of the driving field is $I = 2 \cdot 10^{14}$ W/cm$^2$. In many-atom emission only **(c)** gives a coherent contribution ($\sim N^2$) to the emission, while **(d-f)** give

We next discuss the implications for the emission from many ($N$) atoms. The contributions of the $\boldsymbol{d}_{ii}$ element from Eq. (4) from different atoms add up coherently and the sum is proportional to $N^2$. In contrast, contributions from transitions into states other than the initial state (i.e., the quantum corrections) from different atoms add up incoherently and the sum is proportional to $N$. This effect appears in many areas of physics that involve the combined emission from many atoms (e.g., [28]). For large enough values of $N$, we can eventually neglect the incoherent parts and then the many-atom HHG is adequately captured by Eq. (4) multiplied by $N^2$, in *exact* agreement with the conventional classical theory. However, notably, in our numerical example in Figure 2, the magnitude of $|\boldsymbol{d}_{21}|^2$ (the largest matrix element) is $10^4$ times larger than $|\boldsymbol{d}_{11}|^2$ (Figure 2). Hence, when the number of active atoms is $N < 10^4$, we expect significant observable deviations from the conventional HHG theory Eq. (5), i.e., 'quantum corrections'.

**Section IV – Effects in HHG beyond the dipole approximation**

In all previous works on HHG, the dipole approximation was used for the calculation of the emission. A few papers (e.g., [15]) considered beyond dipole effects for the driving field, but not for the emitted field. In Eqs. (2-5) and in the numerical simulations of Figure 2, we applied the dipole approximation to both the driving field and to the emitted field. In this section, we explore the effects of breaking the dipole approximation for the emitted field. In other words, we derive corrections to the HHG spectrum that result from the extremely short wavelength of the emitted photons themselves (on the same order of or smaller than the size of the electron wavefunction in the driven system).

Firstly, we give an analytical estimate to conditions in which the dipole approximation can be broken. We estimate the effective size of the electron wavefunction during the interaction with the strong driving laser to be on the order of magnitude of the quiver radius $a = \frac{q\lambda_0^2}{4\pi^2 mc^2}\sqrt{\frac{2I}{c\varepsilon_0}}$, where $\lambda_0$ is the wavelength of the driving field. Since $a$ is much smaller than $\lambda_0$, the dipole approximation is accurate for the *driving field*. A typical ratio is $\frac{2\pi}{\lambda_0}a \sim 10^{-2}$ for $\lambda_0 = 800$ nm and $a = 1$ nm. Using a plasmonic environment for confining the field can in principle break the dipole approximation for the driving field as was proposed before in other systems (e.g., [29]), yet plasmonic enhancements of HHG do not currently show evidence of such corrections [30, 31]. We find that the *emitted field* can break the dipole approximation in realistic conditions, which can have subtle implications. The dipole approximation for the emitted field becomes gradually less accurate for higher harmonics, since the wavelength of the emitted field $\lambda$ can reach the single nanometer scale for harmonics on the order $n$ of several hundreds [32].

We find that the HHG emission power scales with the dimensionless parameter $x = \frac{2\pi}{\lambda}a$, or equivalently $x = n \cdot \frac{2\pi}{\lambda_0}a$, so when it becomes on the order of 1, we predict that higher multipolar corrections can become significant. Scenarios in which the emitted field breaks the dipole approximation have not been previously observed nor proposed in the context of HHG. Related effects were predicted before for emission into modes of confined light (e.g., in polaritons in 2D materials [33]), and shown when the emitting electron wavefunction is significantly enlarged (e.g., using Rydberg states [34]). Here, we predict that HHG can break the dipole approximation even with regular atoms emitting into free-space radiation, provided high-enough harmonics, e.g., the 101[th] harmonic [35] causes $x$ to approach unity. Much higher harmonics have been observed [32], and thus by fulfilling the conditions described below, we expect our predictions to be readily observable in existing HHG setups.

To quantify the implications of breaking the dipole approximation, we start from the Schrodinger equation (Eq. (1)) and use exact Hamiltonian without the dipole approximation for the emitted field $H = \frac{1}{2m}\left(\boldsymbol{p} - \frac{q}{c}\boldsymbol{A}\right)^2 + U + H_F$. First order perturbation theory in the weak emitted field leads to the following spectrum:

$$\frac{d\varepsilon}{d\omega d\Omega} = \frac{\omega^2 q^2}{16\pi^3 \varepsilon_0 m^2 c^3} \Sigma_\sigma \left| \int_{-\infty}^{+\infty} \left\langle \phi_i \left| \frac{1}{2}\{(\boldsymbol{p} - q\,\boldsymbol{A}_c(t)), e^{-i\boldsymbol{k}\cdot\boldsymbol{r}}\} \right| \phi_i \right\rangle e^{i\omega t} dt \right|^2, \qquad (6)$$

where $\frac{d\varepsilon}{d\omega d\Omega}$ is the emitted energy per unit solid angle $d\Omega$ per unit frequency $d\omega$, $\{(\boldsymbol{p} - q\,\boldsymbol{A}_c), e^{-i\boldsymbol{k}\cdot\boldsymbol{r}}\} = (\boldsymbol{p} - q\,\boldsymbol{A}_c)e^{-i\boldsymbol{k}\cdot\boldsymbol{r}} + e^{-i\boldsymbol{k}\cdot\boldsymbol{r}}(\hat{\boldsymbol{p}} - q\,\boldsymbol{A}_c)$. We expand Eq. (6) in multipoles and keep the first two terms: the known dipole term $\langle \boldsymbol{d}(\omega) \rangle = \int_{-\infty}^{+\infty} \langle \phi_i(\tau) | q\boldsymbol{r} | \phi_i(\tau) \rangle e^{i\omega\tau} d\tau$ (equivalent to Eq. (5)), and the new quadrupole term $\langle d^2(\omega) \rangle = \int_{-\infty}^{+\infty} \langle \phi_i(\tau) | (q\boldsymbol{r})^2 | \phi_i(\tau) \rangle e^{i\omega\tau} d\tau$ (derivation in SM). To compare the relative contribution of these terms, we extract the length scale of the atomic system from these expressions (see SM) and quantify this dependence using two dimensionless functions $c_1(x)$ and $c_2(x)$ with which we can write:

$$\frac{d\varepsilon}{d\omega} = \frac{1}{8\pi^2 \varepsilon_0 c^3}\left(c_1(x) \cdot \omega^4 |\langle \boldsymbol{d}(\omega)\rangle|^2 + c_2(x) \cdot \frac{1}{(qa)^2}\omega^4 |\langle d^2(\omega)\rangle|^2\right), \qquad (7)$$

where $c_1(x) = \frac{\sin x \cos x - 2x + x \cos 2x + 2x^2 \operatorname{Si}(2x)}{x^3}$, $c_2(x) = \frac{-3x + 4\sin x - \sin x \cos x - 8x \sin^4\frac{x}{2} + 4x^2 \operatorname{Si}(x) - 2x^2 \operatorname{Si}(2x)}{x^3}$.

Importantly, we note that under the dipole approximation for the emitted field ($x \to 0$), we have the following limits $\lim_{x \to 0} c_1(x) = 4/3$ and $\lim_{x \to 0} c_2(x) = 0$ and thus retrieve Eq. (5).

In Figure 3, we calculate the relative contributions of dipolar and quadrupolar emission to the HHG spectrum (dipolar shown in Figure 3a, quadrupolar shown in Figure 3b). While the dipolar emission leads to odd-only harmonics, the quadrupolar emission leads to even-only harmonics. Therefore, even harmonics can be used to measure the quadrupole corrections to HHG. Figure 3c shows the ratio between the quadrupolar and dipolar contributions, by plotting the ratio $c_2(x)/c_1(x)$. The ratio increases with the driving field intensity that increases the size of the electron wavefunction and quiver radius $a$. Moreover, for high harmonics (e.g., soft x-ray photons), quadrupolar contributions become comparable to dipolar contributions. Of course, eventually higher order multipolar corrections can become important too.

Dipolar and quadrupolar emissions also have very different directionality, as can be seen in the insets of Figures 3a and 3b. While dipolar emission is in the direction of propagation of the driving laser, quadrupolar emission has zero intensity in the direction of propagation. When the emission is from many atoms in a large area of interaction relative to the wavelength of the driving laser, the angular distribution of HHG also strongly depends on phase-matching [36] (akin to many other nonlinear processes). Phase-matching in HHG from a single driving laser pulse generally leads the emission to be strongly directional in the direction of driving field propagation. This condition will enhance the dipolar emission and inhibit the quadrupolar emission (which may explain why the breakdown of the dipole approximation has not been observed before). However, in general, the emission direction can be manipulated in various ways. For example, the gas can be confined in a small volume or in a thin layer comparable with the wavelength of the driving field (especially relevant for solid HHG [11, 12, 13, 14, 22]). Alternatively, the driving field can be a superposition of excitations from a few directions with different wave vectors, as done in [37, 38]. In such cases, when the HHG emission is not unidirectional, effects beyond the dipole approximation could play a more important role and change significantly the physics of HHG.

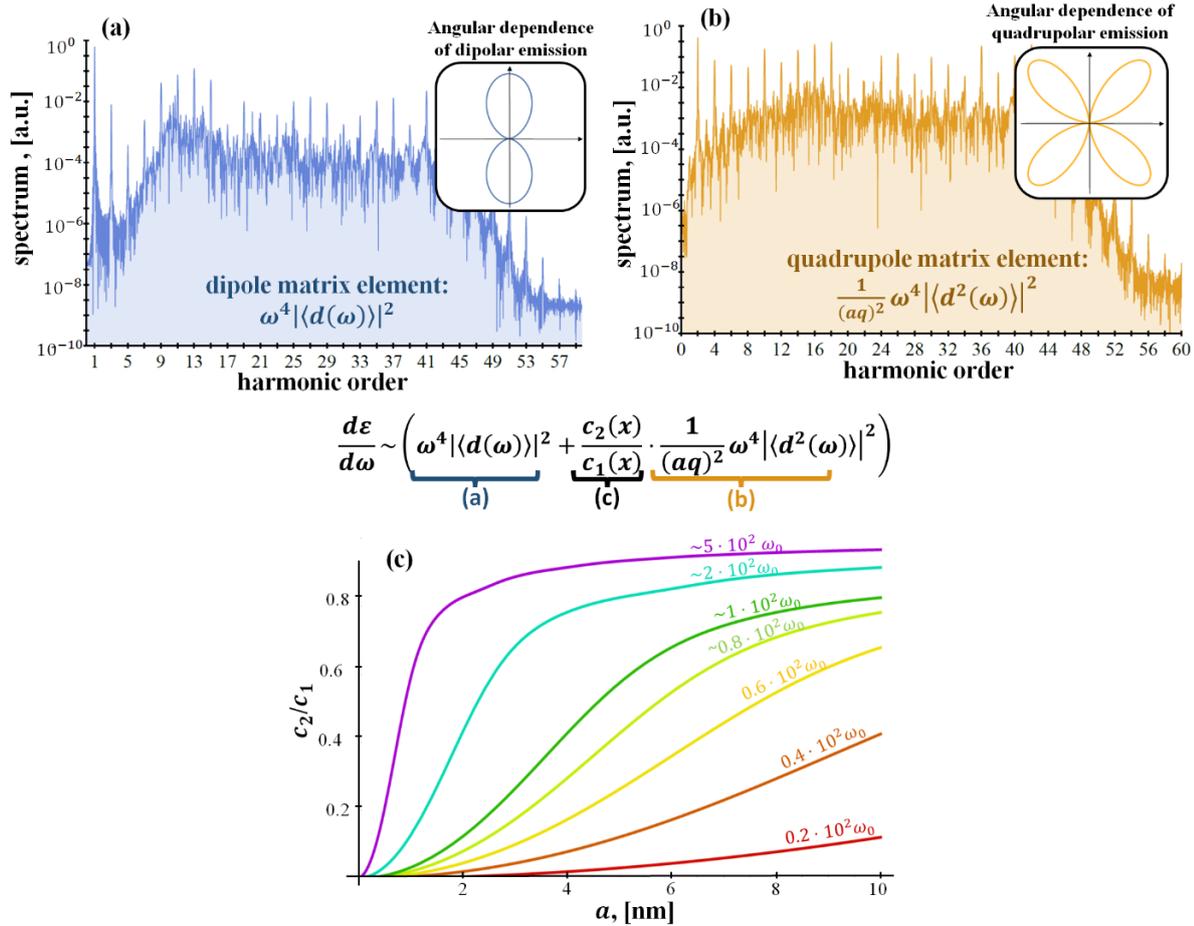

**Figure 3: Breakdown of the dipole approximation for the emitted HHG radiation. (a,b)** Angular dependence and spectrum of dipolar and quadrupolar emission, respectively. Numerical calculations of the spectrum use a 1D model of helium atom. The angular dependence in the insets is derived analytically. The Ox and Oy axes are the direction of propagation and the direction of polarization of the driving field, respectively. The wavelength of the driving field is $\lambda_0 = 800$nm; the intensity of driving field is $I = 2 \cdot 10^{14}$ W/cm². **(c)** Ratio between $c_2$ and $c_1$ from Eq. (7) as a function of the effective size of the electron wavefunction $a$ for different harmonics. This ratio is derived analytically (see SM). The quadrupolar emission increases for higher frequencies and for larger $a$ (which also increases with a stronger driver intensity). The total emission is the weighted sum of **(a)** and **(b)** with the coefficients $c_1, c_2$ from panel **(c)**.

**Section V – Each HHG photon is a comb: discussion of an experimental scheme**

The classical picture of the HHG describes coherent multi-frequency emission that can create an attosecond comb, yet it is not clear from such a picture what is precisely the quantum optical nature of the emission. The extremely nonlinear nature of the process raises many possible descriptions for the quantum nature of the emission, especially since it is constructed from many QED diagrams (Figure 1 right column) in a highly nonperturbative manner. Without the guidance of a complete quantum picture, the emission was for instance considered as being made up from different photons of different frequencies, or as having all photons be of the same frequency in an entangled state with the driving laser. While previous sections dealt with consequences of our SFQED formalism to macroscopic photonic states, below we discuss a consequence on the level of a single photon.

Our formalism reveals the nature of the HHG emission: each emitted photon carries all the frequencies of HHG process. To test this interpretation, we propose the following experiment. Beginning with a typical HHG source (e.g., from a gas of atoms), we can attenuate the output emission to leave on average (less than) one photon per driving laser pulse (Figure 4). We measure the autocorrelation function for the field passing through the attenuator. The autocorrelation result as a function of the attenuator strength could show how the quantum optical nature of HHG scales when the intensity of transmitted field is reduced from a classical field toward the single-photon limit (Figure 4). Our formalism shows that normalized autocorrelation and other measurable quantities will remain the same for *any* number of photons (only up to the challenge of a lower signal-to-noise) (see SM). Such an experiment would lead to a different result for other photonic states and thus could distinguish between them. For example, if the emission is described by classical statistics, normalized autocorrelation function would depend on the number of output photons and in the single photon limit, it would have the form of a cosine function. This experimental setup is very close to the one conducted in [39]. Future work could explore other experimental proposals such as higher order autocorrelation functions (e.g., intensity autocorrelations), and optimize the conditions for such experiments to accommodate for the inevitably low signal in such single-photon experiments.

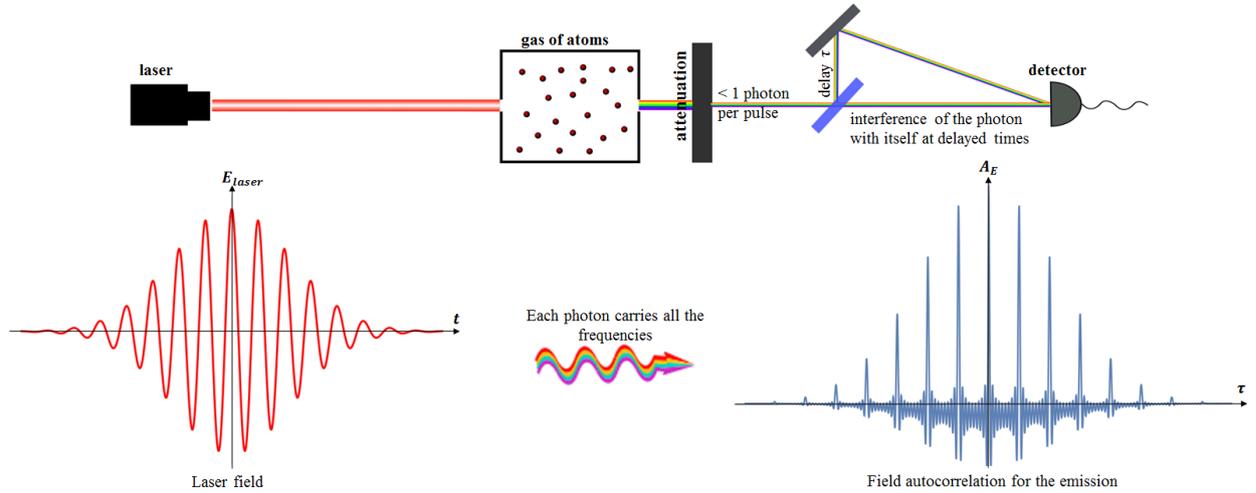

**Figure 4: Proposed autocorrelation experiment: scheme to observe that each HHG is a frequency comb.** HHG emission is first created in typical setup (e.g., from a gas of atoms). We attenuate the output emission and measure the autocorrelation (varying the delay) for different levels of attenuation. If each photon carries the entire spectral content of the HHG emission, then the normalized autocorrelation function would be independent of the number of photons per HHG pulse, even when it is down to less than one photon on average per pulse (see SM).

**Section VI – Summary and Conclusions**

We develop a fully quantum formalism that capture general processes of extreme nonlinear optics, and we demonstrate it for the HHG process. We find new effects that arise from the quantum theory and cannot be described by the conventional theory – effects both on the level of the single HHG photon and on the level of the macroscopic photonic state. Our predictions include the emission of multiple spectral combs, beyond-dipole effects in angle and frequency, and the exact structure of each single photon in HHG.

Our formalism can be straightforwardly generalized in many different directions. We can generalize our theory for systems of many-electrons and for solids, which are the current interest of the community [11, 12, 13, 14, 22]. It is also possible to find the next relativistic corrections to our non-relativistic theory and take in account magnetic dipole effects [40]. Moreover, our formalism is general and relevant not only to the HHG process but to many other processes of nonlinear optics and of quantum electrodynamics as a whole. For instance, the formalism is also capable of reproducing well-known perturbative light-matter interaction effects such as spontaneous emission (see SM). The formalism is applicable for nonlinear Compton Effect [41, 42] and multiphoton Thompson scattering [43].

Looking forward, the results we derive may be of direct importance for the development of novel single-photon sources, such as single-photon frequency-comb sources, which will be of interest in the rising fields of EUV and X-ray quantum optics. Our work may also guide the development of attosecond pulses with quantum features such as entanglement in the extreme ultraviolet or soft-X-ray regime, which may have direct applications for metrology and precision imaging. From a fundamental standpoint, the theory we advance in our manuscript can be applied to all extreme nonlinear optical processes, and thus we expect that our theory will guide the discovery of new quantum effects in other areas of nonlinear optics.